\documentclass{article}
\usepackage{spconf,amsmath,graphicx}
\usepackage{xcolor}
\usepackage{url}
\usepackage{booktabs}
\usepackage[pagebackref,breaklinks,colorlinks]{hyperref}

\title{CO2Wounds-V2: Extended Chronic Wounds Dataset From Leprosy Patients}
\name{%
\begin{tabular}{@{}c@{}}
Karen Sanchez$^{\dagger}$ \qquad 
Carlos Hinojosa$^{\dagger}$ \qquad 
Olinto Mieles$^{\ddagger}$\\ 
Chen Zhao$^{\dagger}$ \qquad 
Bernard Ghanem$^{\dagger}$ \qquad 
Henry Arguello$^{\star}$
\end{tabular}}
 \address{$^{\dagger}$King Abdullah University of Science and Technology, Saudi Arabia.\\
     $^{\ddagger}$Leprosy Control Program, Sanatorio de Contratación ESE, Colombia.\\
     $^{\star}$ Universidad Industrial de Santander, Colombia. \\
{\tt\footnotesize \{karen.sanchez, carlos.hinojosa, chen.zhao, bernard.ghanem\}@kaust.edu.sa} \\ {\tt \footnotesize  mielesburgos@hotmail.com, henarfu@uis.edu.co}}
\begin{document}

\maketitle
\begin{abstract}

Chronic wounds pose an ongoing health concern globally, largely due to the prevalence of conditions such as diabetes and leprosy's disease. The standard method of monitoring these wounds involves visual inspection by healthcare professionals, a practice that could present challenges for patients in remote areas with inadequate transportation and healthcare infrastructure. This has led to the development of algorithms designed for the analysis and follow-up of wound images, which perform image-processing tasks such as classification, detection, and segmentation. However, the effectiveness of these algorithms heavily depends on the availability of comprehensive and varied wound image data, which is usually scarce. This paper introduces the CO2Wounds-V2 dataset, an extended collection of RGB wound images from leprosy patients with their corresponding semantic segmentation annotations, aiming to enhance the development and testing of image-processing algorithms in the medical field.

\end{abstract}
\section{Introduction}
\label{sec:intro}

Chronic wounds impact 40 million individuals worldwide \cite{las2020chronic}. These persistent skin lesions often arise as complications of conditions such as type-2 diabetes, cardiovascular diseases, and neglected tropical illnesses like leprosy, also identified as Hansen's disease \cite{chauhan2003management, van2021psychosocial}. In 2019, the global prevalence of diabetes reached 463 million, with diabetes-related deaths totaling 6.7 million in 2021 \cite{jin2021metabolomics, IDF}. While leprosy cases are fewer in number, the disease remains a significant public health concern, particularly in tropical regions of developing nations. Each year, approximately 225,000 new leprosy cases are reported worldwide, with a notable percentage occurring in children and adolescents \cite{serrano2019social, cardona2018leprosy}. Notably, a considerable proportion of individuals diagnosed with leprosy already exhibit physical impairments at the time of diagnosis, which may become irreversible, contributing to permanent disabling consequences \cite{serrano2019social, cardona2018leprosy}, underlining the enduring impact of this disease.

The treatment of chronic wounds involves regular clinical assessments to monitor for infection, maintain moisture balance, and evaluate wound progression \cite{bowers2020chronic}. However, the management of chronic wounds often faces challenges, leading to potential complications such as limb amputations, infections, and even mortality \cite{yazdanpanah2015literature}. These challenges are further exacerbated in developing countries, where inadequate medical infrastructure and transportation issues hinder access to healthcare facilities, resulting in irregular patient visits and interrupted wound care \cite{gupta2021chronic}.

In recent years, computational methods have emerged as valuable tools to support medical professionals in diagnosing, prognosticating, and treating diseases through automated image processing techniques \cite{calderon2021bilsk, escobar2021accurate}. Specifically, in the domain of chronic wound analysis, image processing algorithms have been employed to detect wound regions and generate segmentation maps, facilitating quantitative assessment of wound characteristics \cite{hsu2019chronic, mukherjee2017diagnostic, chairat2021non}. While existing approaches have demonstrated promising results, limitations persist, including the challenge of adapting models to diverse clinical datasets \cite{sanchez2022cx, goyal2017fully} and limited availability of public data for model training \cite{sanchez2023mask}.

This typical overfitting of deep learning segmentation methods to specific data domains \cite{sanchez2022cx} suggests lower performance when tested on data from different clinical centers or populations, due to substantial differences in wound features, etiology, and image acquisition protocols. 


In this paper, we present the new chronic wounds dataset "CO2Wounds-V2", which contains 764 RGB images of chronic wounds acquired from 96 leprosy patients, with wound semantic segmentation annotations provided in COCO and image formats. This dataset represents the largest collection currently available for chronic wounds caused by leprosy. However, the images in this dataset exhibit variations in appearance, as they were captured using smartphone cameras. Moreover, achieving high accuracy in the segmentation of these wounds poses a significant challenge due to the clinical complexity of these skin lesions. Given the inherent complexity of the dataset in terms of segmentation challenges, it is imperative for the image-processing community to utilize it and devise novel methodologies. Such endeavors will not only aid in overcoming these challenges but also contribute to assisting the medical community in effectively monitoring and managing the treatment of chronic wounds. A full implementation to evaluate the dataset is available at the URL\footnote{\href{https://github.com/simatec-uis/CO2Wounds-V2}{https://github.com/simatec-uis/CO2Wounds-V2}}.

\section{CO2Wounds-V2 Dataset}
\label{sec:dataset}

\subsection{Dataset description}

The CO2Wounds-V2 dataset, publicly available at the URL\footnote{\href{https://data.mendeley.com/datasets/s2w7rjwz49/2}{https://data.mendeley.com/datasets/s2w7rjwz49/2}}, is an updated version of the CO2Wounds database introduced in \cite{monroy2023automated}.
To download, see the “Dataset Download Instructions” section in the supplementary material of this document.
The initial release comprises 164 RGB images of chronic wounds from leprosy patients with their respective detectio labels and segmentation maps. In this paper, we present the latest version, which includes 764 RGB images, divided as follows: 485 for training, 122 for validation, and 157 for testing. The first two sets of this new dataset contain, in addition to the RGB images, their wound annotation labels in COCO format, and binary semantic segmentation masks (wound and background), while the testing split is a collection of unlabeled images.

The images in this dataset were captured by the Leprosy Program control team from the Sanatorio de Contratación ESE in Colombia during routine wound healing sessions. For 1 year and 9 months from November 2021 to August 2023, images were collected from 96 consenting patients including 18 females and 78 males by medical staff, through smartphone cameras. The average age of the participants in this study is 75.5 years old, with a standard deviation of 9.9 years.

\begin{figure*}[t]
    \centering
    \includegraphics[width=.93\linewidth]{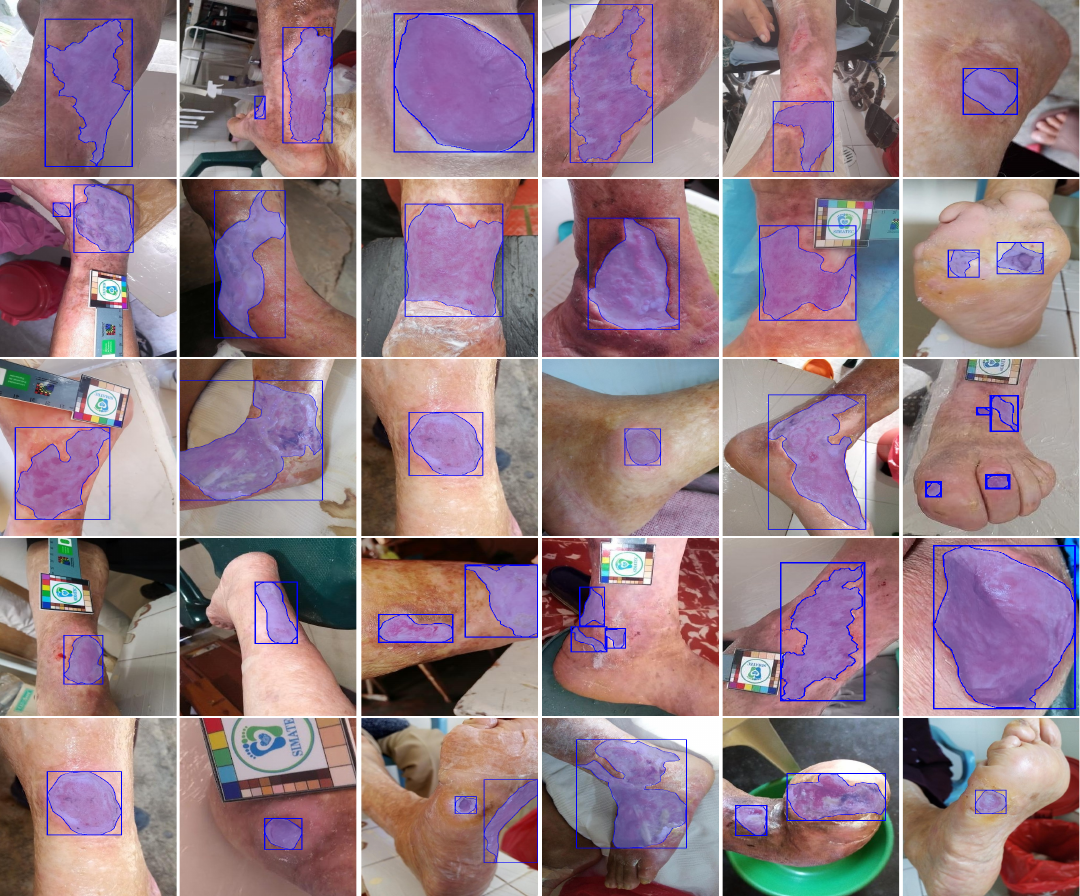}
    \caption{Random image samples from the proposed CO2Wounds-V2 dataset with their annotations. Rectangulars in each image: detection bounding boxes; shaded area in each image: segmentation masks.}
    \label{fig:samples}
\end{figure*}

\subsection{Materials}

\subsubsection{Data collection} Wound images have been acquired using smartphone cameras by medical staff during daily sessions of wound treatment for leprosy patients in a hospital in Colombia.
Each photograph is acquired according to an easily reproducible predefined protocol and submitted to a platform designed by the project team to this end and previously presented in \cite{monroy2023automated}.

The platform is available online on a web page for use by nurses in remote areas to analyze the patient ulcer's condition. It works as a website with a user management system, and each medical staff has access to the platform with a username and password. 
Each user has access to different patients, each with a limited number of wounds associated with them, which can be added to the platform. The wound is the fundamental module of the platform, from within each wound module, medical staff can upload images at different stages and times of the lesion. 

Once an image is uploaded to the platform, it goes through the proposed measurement framework, acquiring the respective segmentation map.  The nurse will then be able to see within the platform a visual comparison of the original and the segmented wound with the metrics of area and perimeter obtained by the algorithm. The individual analyses of the same wound are grouped in a temporal set, to show the evolution of the wound, providing details, and visual and numerical results so that the nurse can calculate if it is improving or progressing to the first image of the wound.

\subsubsection{Data curation} For the dataset creation presented in this paper, the images submitted to the online platform were downloaded while fully preserving patient privacy, without including any patient identity information. Then, a  data curation step was developed. This step included deleting duplicated images, images that do not contain wounds, and blurring images where the edges of the ulcer are difficult to detect. After that, 746 photographs remained, and all of them were then cropped to a 4:3 aspect ratio.

\subsubsection{Annotation} With the curated 746 images, we meticulously delineated the wound outlines on 607 of them through the Computer Vision Annotation Tool\footnote{\href{www.cvat.ai}{cvat.ai}} to form the training and validation sets. Semantic wound segmentation annotations are provided in COCO and image formats, all of which are included in this dataset.  We demonstrate some annotated examples in Fig.~\ref{fig:samples}.

\subsection{Ethical aspects}

This dataset was acquired in accordance with the basic principles of Autonomy, Beneficence, Non-maleficence, Justice, and Respect.

\subsubsection{Autonomy} Patients were guaranteed the right to decide freely, informed of the facts and without any kind of coercion, about whether they were willing to participate in the dataset acquisition project. This principle was specified with the signing of an \textit{informed consent}, designed following articles 15 and 16 of Resolution 8430 of 1993 of the Colombian Law. If the patient did not wish to participate in the project, her wounds would not be photographed, and this would not affect her care or the quality of her treatment in any way.

\subsubsection{Beneficence} Health professionals acquire photographs of chronic ulcers present in the lower extremities, focusing only on the area in which the ulcer occurs, without photographing the patient's face at any time. The information is anonymous and confidential, safeguarding the patient's identity.

\subsubsection{Non-maleficence} The researchers declare that they did not acquire, process, or share information without the prior consent of the patients; furthermore, under no circumstances will the identity of the participants be disclosed, and no actions will be taken that would cause harm or damage to anyone involved in the project, nor will any additional risks beyond those considered in the ethical aspects of the research be pursued.

\subsubsection{Justice} 
The present project respects and guarantees the national policy for the treatment of personal data established in Statutory Law 1581 of 2012, its Regulatory Decree 1377 of 2013, and institutional Rector's Resolution 1227 of 2013 from the Universidad Industrial de Santander. The patients are the owners of their information, they have all the rights regarding their data, including free access, the possibility of rectification, ownership, the decision to withdraw, and free abstention, among others.

\subsubsection{Respect} 
In recognition of the right to respect patient dignity and the tasks inherent to each profession, the computer science personnel associated with this research project did not have knowledge of, manipulate, or have direct contact with patients. Their involvement was limited to interactions with medical staff and the handling of anonymized photographs of lower limbs for data curation and preparation.

\subsection{License} The authors make data publicly available according to open data standards and license datasets under the Creative Commons Attribution-NonCommercial-NoDerivatives (CC BY-NC-ND) license\footnote{\href{https://doi.org/10.17632/s2w7rjwz49.2}{https://doi.org/10.17632/s2w7rjwz49.2}}.

\section{Evaluation}

\subsection{Metrics}
\label{metrics}

To evaluate the performance of the semantic segmentation models, we report standard metrics widely used in the computer vision community. These include the \textit{mean Intersection over Union (mIoU)}, which quantifies the pixel-wise overlap between the predicted segmentation maps and the ground truth \cite{everingham2015pascal}. The \textit{F1 score}, a harmonic mean of precision and recall, offers a balance between the accuracy of the model in identifying relevant pixels and its ability to minimize false positives. \textit{Accuracy} measures the proportion of correctly classified pixels across the entire dataset. \textit{Precision} assesses the ability of the model to accurately predict positive pixels, while \textit{recall} focuses on the capability of the model to identify all relevant pixels in the ground truth. Collectively, these metrics provide a comprehensive assessment of the performance of the semantic segmentation models tested with our CO2Wounds-V2 dataset. The reported metrics were computed using the implementation provided by SMP library \cite{Iakubovskii:2019}.

For future research using this dataset, we recommend providing the same standard semantic segmentation metrics so that comparison with other works addressing the same task is possible. 



\subsection{Baseline Performance}

\begin{figure}[t!]
    \centering
    \includegraphics[width=\columnwidth]{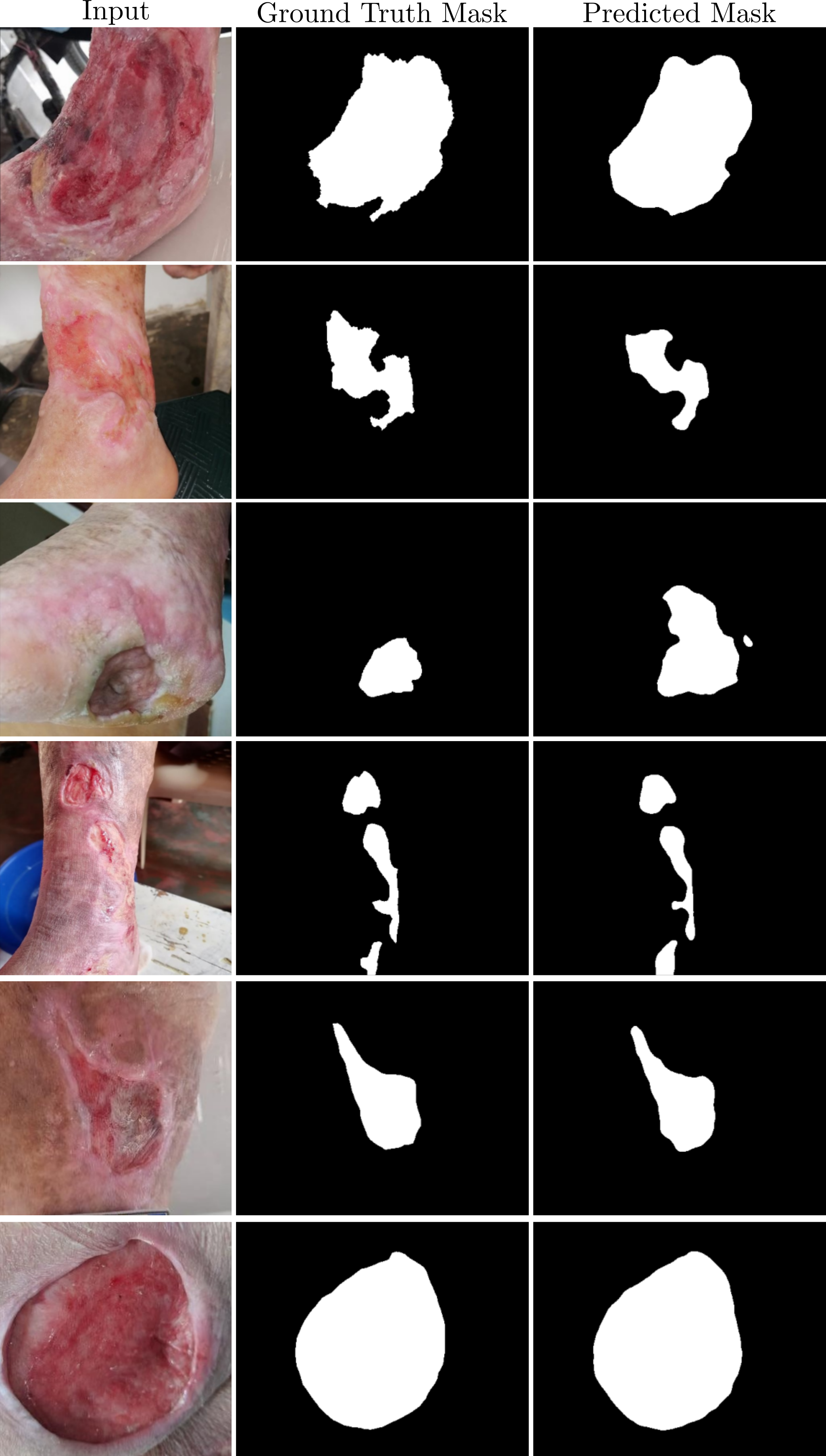}
    \caption{Random samples from the validation set of our CO2Wounds-V2 dataset (Column 1), ground truth segmentation masks (Column 2), and segmentation results using U-Net architecture (Column 3).}
    \label{fig:results}
\end{figure}

In this section, we conduct an initial semantic segmentation trial using the CO2Wounds-V2 dataset to establish a baseline for future experiments. Our evaluation involved exploring different configurations of deep learning convolutional neural networks (CNN). For this purpose, we use PyTorch \cite{paszke2019pytorch} alongside the Segmentation Models PyTorch Library (SMP) \cite{Iakubovskii:2019} for performing wound semantic segmentation on our dataset. SMP offers a collection of deep neural network models specifically designed for segmentation tasks, all of which are built on top of PyTorch. 
Specifically, we trained three different well-known semantic segmentation architectures: Feature Pyramid Network (FPN) \cite{kirillov2019panoptic}, U-Net \cite{ronneberger2015u}, and DeepLabV3+ \cite{chen2018encoder}. All the networks were trained during 100 epochs using the Adam optimizer.
Different encoders with Imagenet pre-trained weights were used to evaluate the architecture's performance on our proposed dataset, including ResNeXt-50, ResNet-101, EfficientNet, and SegFormer. Table \ref{tab:results} gives an overview of the results with the metrics described in section \ref{metrics}.

\begin{table*}[t]
\caption{Performance comparison of different segmentation architectures on our proposed CO2Wounds-V2 dataset.}
\centering
\begin{tabular}{@{}ccccccc@{}}
\toprule
\textbf{Architectures} & \textbf{Encoder} & \textbf{mIoU(\%)}  & \textbf{F1(\%)}    & \textbf{Accuracy(\%)} & \textbf{Precision(\%)} & \textbf{Recall(\%)} \\ \midrule
DeepLabV3              & ResNeXt-50       & 68.48          & 77.86          & 98.58             & \textbf{85.81}              & 78.20           \\
DeepLabV3+             & ResNeXt-50       & 68.23          & 78.04          & 98.51             & 81.61              & 81.69           \\
U-Net                  & ResNeXt-50       & 69.94          & \textbf{79.44} & 98.65             & 84.35              & 80.31           \\
FPN                    & ResNeXt-50       & 68.99          & 78.36          & 98.53             & 82.94              & 81.17           \\ \midrule
DeepLabV3+             & ResNet-101       & 66.88          & 76.55          & 98.40             & 83.04              & 79.02           \\
U-Net                  & ResNet-101       & 66.96          & 76.61          & 98.28             & 80.87              & 81.10           \\
FPN                    & ResNet-101       & 66.81          & 76.52          & 98.45             & 81.78              & 79.61           \\ \midrule
DeepLabV3+             & EfficientNet     & 66.98          & 76.78          & 98.49             & 79.88              & 81.86           \\
U-Net                  & EfficientNet     & 67.71          & 77.20          & 98.51             & 83.29              & 77.80           \\
FPN                    & EfficientNet     & 67.49          & 76.84          & \textbf{98.59}             & 82.63              & 80.34           \\ \midrule
U-Net                  & SegFormer        & \textbf{70.13} & 79.26          & \textbf{98.59}             & 84.70              & 81.99           \\
FPN                    & SegFormer        & 69.90          & 79.36          & 98.56             & 82.02              & \textbf{84.35}           \\ \bottomrule
\end{tabular}
\label{tab:results}
\end{table*}


\subsection{Results and Discussions}

In Figure \ref{fig:results}, we present six randomly selected samples from the validation set of our CO2Wound-V2 dataset. Each sample is accompanied by its corresponding ground truth segmentation map and the segmentation map is predicted through the U-Net architecture, which depicted the highest mIoU performance and F1-score in our baseline experiments. While we have achieved good visual results in our baseline experiments, we acknowledge that there is room for improvement. As this paper primarily focuses on introducing and providing a benchmark dataset, we anticipate that future researchers will build upon our work to further enhance the segmentation results. We encourage researchers to explore the dataset and contribute to the advancement of medical image analysis techniques for wound assessment and treatment monitoring.

\section{Conclusions}

We have introduced the CO2Wounds-V2 dataset, which contributes to the automated wound analysis techniques. By providing 764 RGB images, divided into training, validation, and testing sets, along with corresponding annotations.
We offer a comprehensive benchmark for evaluating the performance of segmentation algorithms. The dataset captures a diverse range of chronic wounds from leprosy patients, collected over 1 year and 9 months in a clinical setting of a developing country. Notably, the inclusion of images captured using smartphone cameras adds to the dataset's real-world applicability. While our baseline experiments have shown promising results, we anticipate that the research community will leverage this dataset to develop novel algorithms and further improve the state-of-the-art in automated wound analysis.

\section{Compliance with ethical standards}

The acquisition and use of the CO2Wounds-V2 database were performed in line with the Declaration of Helsinki principles. The corresponding approval was granted by the Ethics Committee of the Sanatorio de Contratación ESE hospital in Act 05–21, and also by the Scientific Research Ethics Committee of the Universidad Industrial de Santander in Act No. 19-13/11/20.

\section{Acknowledgments}
We thank Brayan Monroy, Paula Arguello, Claudia V Correa, Juan Estupiñán, Jorge Bacca, and Fernando Rojas for their contribution to the development of the \textit{Simatec} project and to the creation of the online platform used to collect this data.
This work was supported by the Vicerrectoría de Investigación y Extensión of the Universidad Industrial de Santander, Colombia, under the research project 2707. This work was supported by the SDAIA-KAUST Center of Excellence in Data Science and Artificial Intelligence (SDAIA-KAUST AI) under the project ID RGC/3/4816-06-01.

\bibliographystyle{IEEEbib}
\bibliography{refs}

\begin{thebibliography}{10}

\bibitem{las2020chronic}
Kevin Las~Heras, Manoli Igartua, Edorta Santos-Vizcaino, and Rosa~Maria Hernandez,
\newblock ``Chronic wounds: Current status, available strategies and emerging therapeutic solutions,''
\newblock {\em Journal of Controlled Release}, vol. 328, pp. 532--550, 2020.

\bibitem{chauhan2003management}
VS~Chauhan, SS~Pandey, and VK~Shukla,
\newblock ``Management of plantar ulcers in hansen's disease,''
\newblock {\em The international journal of lower extremity wounds}, vol. 2, no. 3, pp. 164--167, 2003.

\bibitem{van2021psychosocial}
Robin van Wijk, Lena van Selm, Martha~C Barbosa, Wim~H van Brakel, Mitzi Waltz, and Karl~Philipp Puchner,
\newblock ``Psychosocial burden of neglected tropical diseases in eastern colombia: an explorative qualitative study in persons affected by leprosy, cutaneous leishmaniasis and chagas disease,''
\newblock {\em Global Mental Health}, vol. 8, 2021.

\bibitem{jin2021metabolomics}
Qiao Jin and Ronald Ching~Wan Ma,
\newblock ``Metabolomics in diabetes and diabetic complications: Insights from epidemiological studies,''
\newblock {\em Cells}, vol. 10, no. 11, pp. 2832, 2021.

\bibitem{IDF}
{International Diabetes Federation},
\newblock ``{IDF} diabetes atlas, 10th edn. {Brussels, Belgium}: 2021. {Available at: \url{https://www.diabetesatlas.org}},'' 2021,
\newblock Accessed: 2021-07-12.

\bibitem{serrano2019social}
H{\'e}ctor Serrano-Coll, Hugo~Rene Mora, Juan~Camilo Beltr{\'a}n, Malcolm~S Duthie, and Nora Cardona-Castro,
\newblock ``Social and environmental conditions related to mycobacterium leprae infection in children and adolescents from three leprosy endemic regions of colombia,''
\newblock {\em BMC infectious diseases}, vol. 19, no. 1, pp. 1--10, 2019.

\bibitem{cardona2018leprosy}
Nora Cardona-Castro,
\newblock ``Leprosy in colombia,''
\newblock {\em Current Tropical Medicine Reports}, vol. 5, no. 2, pp. 85--90, 2018.

\bibitem{bowers2020chronic}
Steven Bowers and Eginia Franco,
\newblock ``Chronic wounds: evaluation and management,''
\newblock {\em American family physician}, vol. 101, no. 3, pp. 159--166, 2020.

\bibitem{yazdanpanah2015literature}
Leila Yazdanpanah, Morteza Nasiri, and Sara Adarvishi,
\newblock ``Literature review on the management of diabetic foot ulcer,''
\newblock {\em World journal of diabetes}, vol. 6, no. 1, pp. 37, 2015.

\bibitem{gupta2021chronic}
Shivani Gupta, Sushma Sagar, Girisha Maheshwari, Tomohika Kisaka, and Suteerth Tripathi,
\newblock ``Chronic wounds: Magnitude, socioeconomic burden and consequences,''
\newblock {\em Wounds Asia}, vol. 4, pp. 8--14, 2021.

\bibitem{calderon2021bilsk}
Camilo Calder{\'o}n, Karen Sanchez, Sergio Castillo, and Henry Arguello,
\newblock ``Bilsk: A bilinear convolutional neural network approach for skin lesion classification,''
\newblock {\em Computer Methods and Programs in Biomedicine Update}, vol. 1, pp. 100036, 2021.

\bibitem{escobar2021accurate}
Jessica Escobar, Karen Sanchez, Carlos Hinojosa, Henry Arguello, and Sergio Castillo,
\newblock ``Accurate deep learning-based gastrointestinal disease classification via transfer learning strategy,''
\newblock in {\em 2021 XXIII Symposium on Image, Signal Processing and Artificial Vision (STSIVA)}. IEEE, 2021, pp. 1--5.

\bibitem{hsu2019chronic}
Jui-Tse Hsu, Yung-Wei Chen, Te-Wei Ho, Hao-Chih Tai, Jin-Ming Wu, Hsin-Yun Sun, Chi-Sheng Hung, Yi-Chong Zeng, Sy-Yen Kuo, and Feipei Lai,
\newblock ``Chronic wound assessment and infection detection method,''
\newblock {\em BMC medical informatics and decision making}, vol. 19, no. 1, pp. 1--20, 2019.

\bibitem{mukherjee2017diagnostic}
Rashmi Mukherjee, Suman Tewary, and Aurobinda Routray,
\newblock ``Diagnostic and prognostic utility of non-invasive multimodal imaging in chronic wound monitoring: a systematic review,''
\newblock {\em Journal of medical systems}, vol. 41, no. 3, pp. 1--17, 2017.

\bibitem{chairat2021non}
Sawrawit Chairat, Tulaya Dissaneewate, Piyanun Wangkulangkul, Laliphat Kongpanichakul, and Sitthichok Chaichulee,
\newblock ``Non-contact chronic wound analysis using deep learning,''
\newblock in {\em 2021 13th Biomedical Engineering International Conference (BMEiCON)}. IEEE, 2021, pp. 1--5.

\bibitem{sanchez2022cx}
Karen Sanchez, Carlos Hinojosa, Henry Arguello, Denis Kouam{\'e}, Olivier Meyrignac, and Adrian Basarab,
\newblock ``Cx-dagan: Domain adaptation for pneumonia diagnosis on a small chest x-ray dataset,''
\newblock {\em IEEE Transactions on Medical Imaging}, 2022.

\bibitem{goyal2017fully}
Manu Goyal, Moi~Hoon Yap, Neil~D Reeves, Satyan Rajbhandari, and Jennifer Spragg,
\newblock ``Fully convolutional networks for diabetic foot ulcer segmentation,''
\newblock in {\em 2017 IEEE international conference on systems, man, and cybernetics (SMC)}. IEEE, 2017, pp. 618--623.

\bibitem{sanchez2023mask}
Karen Sanchez, Carlos Hinojosa, Kevin Arias, Henry Arguello, Denis Kouam{\'e}, Olivier Meyrignac, and Adrian Basarab,
\newblock ``Mask-guided data augmentation for multiparametric mri generation with a rare hepatocellular carcinoma,''
\newblock in {\em 2023 IEEE 20th International Symposium on Biomedical Imaging (ISBI)}. IEEE, 2023, pp. 1--5.

\bibitem{monroy2023automated}
Brayan Monroy, Karen Sanchez, Paula Arguello, Juan Estupi{\~n}{\'a}n, Jorge Bacca, Claudia~V Correa, Laura Valencia, Juan~C Castillo, Olinto Mieles, Henry Arguello, et~al.,
\newblock ``Automated chronic wounds medical assessment and tracking framework based on deep learning,''
\newblock {\em Computers in Biology and Medicine}, vol. 165, pp. 107335, 2023.

\bibitem{everingham2015pascal}
Mark Everingham, SM~Ali Eslami, Luc Van~Gool, Christopher~KI Williams, John Winn, and Andrew Zisserman,
\newblock ``The pascal visual object classes challenge: A retrospective,''
\newblock {\em International journal of computer vision}, vol. 111, pp. 98--136, 2015.

\bibitem{Iakubovskii:2019}
Pavel Iakubovskii,
\newblock ``Segmentation models pytorch,'' \url{https://github.com/qubvel/segmentation_models.pytorch}, 2019.

\bibitem{paszke2019pytorch}
Adam Paszke, Sam Gross, Francisco Massa, Adam Lerer, James Bradbury, Gregory Chanan, Trevor Killeen, Zeming Lin, Natalia Gimelshein, Luca Antiga, et~al.,
\newblock ``Pytorch: An imperative style, high-performance deep learning library,''
\newblock {\em Advances in neural information processing systems}, vol. 32, 2019.

\bibitem{kirillov2019panoptic}
Alexander Kirillov, Ross Girshick, Kaiming He, and Piotr Doll{\'a}r,
\newblock ``Panoptic feature pyramid networks,''
\newblock in {\em Proceedings of the IEEE/CVF conference on computer vision and pattern recognition}, 2019, pp. 6399--6408.

\bibitem{ronneberger2015u}
Olaf Ronneberger, Philipp Fischer, and Thomas Brox,
\newblock ``U-net: Convolutional networks for biomedical image segmentation,''
\newblock in {\em Medical Image Computing and Computer-Assisted Intervention--MICCAI 2015: 18th International Conference, Munich, Germany, October 5-9, 2015, Proceedings, Part III 18}. Springer, 2015, pp. 234--241.

\bibitem{chen2018encoder}
Liang-Chieh Chen, Yukun Zhu, George Papandreou, Florian Schroff, and Hartwig Adam,
\newblock ``Encoder-decoder with atrous separable convolution for semantic image segmentation,''
\newblock in {\em Proceedings of the European conference on computer vision (ECCV)}, 2018, pp. 801--818.

\end{thebibliography}

\end{document}